# Subharmonic oscillations in the Floquet circuit with the frequency synthesis dimension

Bo Lv[1*], Shiyun Xia[1], Ye Tian[2], Ting Liu[3], Hongyang Mu[1], Zhichao Shen[1], Sijie Wang[1], Zheng Zhu[1], Huibin Tao[4], Fanyi Meng[5*], Jinhui Shi[1*]

[1]Key Laboratory of In-Fiber Integrated Optics of Ministry of Education, College of Physics and Optoelectronic Engineering, Harbin Engineering University, Harbin 150001, Heilongjiang Province, China

[2]College of Automation, Nanjing University of Science and Technology, Nanjing 210000, Jiangsu Province, China

[3]College of Underwater Acoustic Engineering, Harbin Engineering University, Harbin 150001, China

[4]School of Software Engineering, Xi'an Jiaotong University, Xi'an, China

[5]Department of microwave engineering, school of electronics and information, Harbin Institute of Technology, Harbin 150001, Heilongjiang Province, China

*Correspondence to:

bolv@hrbeu.edu.cn, blade@hit.edu.cn, shijinhui@hrbeu.edu.cn.

**Abstract**

The period-doubling oscillation emerges due to the coexistence of zero and π modes in Floquet topological insulators (FTIs). Here, leveraging the flexibility of the circuit, we construct a classical circuit with frequency synthetic dimension to realize the Floquet Hamiltonian of a periodically-driven model and demonstrate the topological edge states of zero and π modes. In contrast to the period-doubling oscillations observed in FTIs, the circuit exhibits deeply-subharmonic oscillations with periods extensively exceeding the doubling-driven period. Furthermore, we explore the band of the circuit with the equivalent-enhanced periodically-driven strength. Our method provides a flexible scheme to study Floquet topological phases, and open a new path for realizing the deeply subwavelength system.

**Key words**: Classical circuit, Floquet topological phases, deeply-subharmonic oscillation, frequency-synthesis dimension.

## Introduction

In recent years, topological insulators (TIs) have attracted numerous investigation due to their unique physical properties [1-5]. Within this domain, Floquet topological insulators (FTIs), characterized by their response to periodic modulation [6-10], exhibit many distinctive topological phenomena, including topological $\pi$ modes [11-12], double-periodic resonance [13-14], Floquet-Majorana states [15-18] and combining non-Hermitian systems [19], quantum Hall conductors [20, 21], high-dimensional [22-27] and high-order topological insulators [28-37]. Various theoretical and experimental platforms have been proposed for realization of FTIs. In the realm of condensed matter physics, investigations into Floquet topological phases have been conducted using ultracold atom lattices [38], superconductors [40-50] and semiconductor quantum wells [51]. Meanwhile, in the context of classical physics, Floquet topological phases have been realized within optical [52-57] and acoustic systems [58-59], demonstrating the broad applicability and interdisciplinary interest in these driven steady states of matter.

Due to the challenges associated with the realization of time regulation in condensed matter, optical and acoustic systems, researchers have tuned to alternative methods to realize FTIs. The mechanism of periodic modulation is achieved by constructing helicity waveguide or employing periodically varied-spatial dimension [6, 11, 13, 59]. However, these approaches are subject to spatial-geometrical constraints, which complicate assembly and the range of modulable-frequencies. Among classical physical systems, electric circuits with their highly flexible performance serve as suitable platforms of FTIs [60-64]. Here, periodic modulation can be implemented through the use of active elements or spatial modulation [65-67].

In this study, the Floquet Hamiltonian of the periodically-driven model is facilitated by the introduction of a frequency-synthetic dimension, achieved by implementing the circuit-oscillator hierarchy with the stepping-variation resonances. The architecture of the circuit incorporates a chain in each row, braided according to the Su-Schrieffer-Heeger (SSH) model, where the on-site resonators consist of grounded in-parallel inductors and capacitors, and the hopping terms are implemented by intralayer-coupling capacitors. Adjacent rows of resonators are interconnected via interlayer-coupling capacitors. When the difference in resonant frequency between adjacent rows decreases below the static band of the SSH circuit chain, topological π modes manifest within the band. Subsequent adjustment of the values of the intralayer-coupling capacitors leads to the emergence of zero modes. The coexistence of π and zero modes induces deeply subharmonic oscillations at the edge of the circuit. Furthermore, we explore the band behavior under conditions of significant time-modulation amplitude. The developed circuit holds potential for applications in generating very low frequency (VLF) signals, and moreover, it provides a novel method for achieving deeply long-wavelength resonance in various systems.

**The construction of the circuit**

The Hamiltonian with time-periodic evolution $H(t+T)=H(t)$ in momentum space reads

$$H(k,t)=\begin{pmatrix} 0 & \tilde{t}_a(t)+\tilde{t}_b(t)e^{-ik} \\ \tilde{t}_a(t)+\tilde{t}_b(t)e^{ik} & 0 \end{pmatrix}, \tag{1}$$

where $\tilde{t}_{a/b}(t)=t_{a/b}+2V\cos(\Omega t)$ represent the time-dependent parameters with the driven frequency $\Omega=\frac{2\pi}{T}$, and $t_{a/b}$ represent the static terms of $\tilde{t}_{a/b}(t)$. The time-dependent dynamic equation has the form:

$$\left(H(k,t)-i\partial_t\right)\left|\psi(t)\right\rangle=0, \tag{2}$$

where $H_F=H(k,t)-i\partial_t$ is the quasienergy operator [68]. The Floquet state can be expressed as $\left|\psi(t)\right\rangle=\exp(-i\varepsilon t)\left|\varphi(t)\right\rangle$, where $\varepsilon$ and $\left|\varphi(t)\right\rangle$ are the quasienergy and time-periodic mode, respectively. the Floquet states are expanded by Fourier series for getting rid of the time-dependent part. The Fourier expansion of the Floquet state at the quasienergy $\varepsilon_n$ has the form:

$$\left|\psi_n(t)\right\rangle=\exp(-i\varepsilon_n t)\sum_m c_m\exp(-im\Omega t)\left|\phi_n^{(m)}\right\rangle, \tag{3}$$

where $\left|\phi_n^{(m)}\right\rangle$ is the *m*-th Fourier component of $\left|\varphi_n(t)\right\rangle$, and $c_m$ is the superposition coefficients. In the Fourier space, the equation of the quasienergy $\varepsilon_n$ reads:

$$\sum_{m'} H_{mm'} \left|\phi_n^{(m')}\right\rangle = \varepsilon_n \left|\phi_n^{(m)}\right\rangle, \tag{4}$$

where the Hamiltonian in Fourier space is expanded as $H_{mm'} = m\Omega\delta_{mm'}\mathbf{I} + H_{m-m'}$, in which $H_{m-m'}$ is the Fourier component of the time-dependent Hamiltonian $H(t)$ and has the form $H_l = \frac{1}{T}\int_0^T \exp(-il\Omega t) H(t) dt$ with the index $l = m - m'$. The matrix of $H_F$ has form:

$$H_F = \begin{pmatrix} \ddots & \vdots & \vdots & \vdots & \vdots & \vdots & \iddots \\ \cdots & H_0\text{-}2\Omega & H_{-1} & 0 & 0 & 0 & \cdots \\ \cdots & H_1 & H_0\text{-}\Omega & H_{-1} & 0 & 0 & \cdots \\ \cdots & 0 & H_1 & H_0 & H_{-1} & 0 & \cdots \\ \cdots & 0 & 0 & H_1 & H_0\text{+}\Omega & H_{-1} & \cdots \\ \cdots & 0 & 0 & 0 & H_1 & H_0 + 2\Omega & \cdots \\ \iddots & \vdots & \vdots & \vdots & \vdots & \vdots & \ddots \end{pmatrix}, \tag{5}$$

where the Fourier components of the Hamiltonian in Eq. 1 have the forms:

$$H_0 = \begin{pmatrix} 0 & t_a + t_b e^{-ik} \\ t_a + t_b e^{ik} & 0 \end{pmatrix}, \tag{6}$$

$$H_1 = H_{-1} = \begin{pmatrix} 0 & V\left(1 + e^{-ik}\right) \\ V\left(1 + e^{ik}\right) & 0 \end{pmatrix}, \tag{7}$$

and $H_l = 0$ for $l \geqslant 2$. Based on above analysis, the matrix of the $H_F$ is expressed:

$$H_F = \begin{pmatrix} \ddots & \vdots & \vdots & \vdots & \vdots & \vdots & \vdots & \iddots \\ \cdots & -\Omega & t_a + t_b e^{-ik} & 0 & V\left(1+e^{-ik}\right) & 0 & 0 & \cdots \\ \cdots & t_a + t_b e^{ik} & -\Omega & V\left(1+e^{ik}\right) & 0 & 0 & 0 & \cdots \\ \cdots & 0 & V\left(1+e^{-ik}\right) & 0 & t_a + t_b e^{-ik} & 0 & V\left(1+e^{-ik}\right) & \cdots \\ \cdots & V\left(1+e^{ik}\right) & 0 & t_a + t_b e^{ik} & 0 & V\left(1+e^{ik}\right) & 0 & \cdots \\ \cdots & 0 & 0 & 0 & V\left(1+e^{-ik}\right) & \Omega & t_a + t_b e^{-ik} & \cdots \\ \cdots & 0 & 0 & V\left(1+e^{ik}\right) & 0 & t_a + t_b e^{ik} & \Omega & \cdots \\ \iddots & \vdots & \vdots & \vdots & \vdots & \vdots & \vdots & \ddots \end{pmatrix}. \tag{8}$$

According to the schematic of $H_F$, we construct the circuit as shown in Fig. 1a, and the components of the model are listed in Fig. 1b. The label of the row is indicated as the order $m$ as $\{m \mid m \in [-M, M] \text{ and } m \in \mathbb{Z}\}$, and total number of the row is $2M + 1$.

The circuit chain in each row is designed as the SSH model, with the on-site section consisting of the grounded-parallel inductor $L$ and capacitor $C_m$ (referred to as grounded-$LC_m$ resonator), and the intralayer-hopping terms implemented by the coupling capacitors $C_a$ and $C_b$. The values of the grounded capacitors in different rows exhibit stepped variation with $C_{m+1} - C_m = C_\Omega$, and the nodes between the nearest-neighbor rows are connected by the interlayer-capacitor $C_v$. Here the circuit model consists of five rows, with the maximum order is $M = 2$. In contrast to the Floquet Hamiltonian of the periodically-driven system with infinite copies of bands, the circuit with finite chains can also exhibit Floquet topological edge modes. The analog models of Floquet Hamiltonian's construction are implemented in literatures 65, 69 and 70. According to Bloch's theorem, the dynamic equation of the circuit with infinite boundary in $k$ space is calculated as [60-64]

$$\frac{\omega_0^2}{\omega^2} V = \mathcal{H} V, \tag{9}$$

where $\omega_0 = \frac{1}{\sqrt{LC_0}}$ is the resonant frequency of the grounded-$LC_0$ resonator in the row of the order $m = 0$. $\omega$ is the frequency of the signal on the circuit. The Hamiltonian $\mathcal{H}(k)$ in Eq. 9 possesses the form:

$$\mathcal{H}(k) = \begin{pmatrix}
t_{on}-2t_\Omega & -t_a-t_b e^{-ik} & & -t_v(1+e^{-ik}) & & & & & & \\
-t_a-t_b e^{ik} & t_{on}-2t_\Omega & -t_v(1+e^{ik}) & & & & & & & \\
 & -t_v(1+e^{-ik}) & t_{on}-t_\Omega & -t_a-t_b e^{-ik} & & -t_v(1+e^{-ik}) & & & & \\
-t_v(1+e^{ik}) & & -t_a-t_b e^{ik} & t_{on}-t_\Omega & -t_v(1+e^{ik}) & & & & & \\
 & & & -t_v(1+e^{-ik}) & t_{on} & -t_a-t_b e^{-ik} & & -t_v(1+e^{-ik}) & & \\
 & & -t_v(1+e^{ik}) & & -t_a-t_b e^{ik} & t_{on} & -t_v(1+e^{ik}) & & & \\
 & & & & & -t_v(1+e^{-ik}) & t_{on}+t_\Omega & -t_a-t_b e^{-ik} & & -t_v(1+e^{-ik}) \\
 & & & & -t_v(1+e^{ik}) & & -t_a-t_b e^{ik} & t_{on}+t_\Omega & -t_v(1+e^{ik}) & \\
 & & & & & & & -t_v(1+e^{-ik}) & t_{on}+2t_\Omega & -t_a-t_b e^{-ik} \\
 & & & & & & -t_v(1+e^{ik}) & & -t_a-t_b e^{ik} & t_{on}+2t_\Omega
\end{pmatrix}, \tag{10}$$

where the parameters $t_\Omega = \frac{C_\Omega}{C_0}$, $t_v = \frac{C_v}{C_0}$ and $t_{a/b} = \frac{C_{a/b}}{C_0}$ are the capacitance coefficients in the circuit, and $t_{on}$ in the diagonal elements is $t_{on} = 1 + t_a + t_b + 4t_v$. The eigenvalues of the Hamiltonian in Eq. 10 correspond to the square of the normalized resonant frequencies of the circuit. Comparing with the form of $\mathcal{H}(k)$, the capacitance parameters in Eq. 10 embody the correspondence relations as $t_\Omega \leftrightarrow \Omega$, $t_v \leftrightarrow V$, where $V$ presents the amplitude of the driven frequency of the time modulation here, and indicates the voltage of the circuit in the following content. Furthermore, the intralayer-coupling parameters in Eq. 8 and Eq. 10 share the same labels $t_{a/b}$. In Eq. 10, the parameters $t_\Omega$ and $t_v$ are named as the Driven Frequency (DF) capacitor and the Amplitude of the Periodic Modulation (APM) capacitor, respectively. Although the on-site energies of the Floquet Hamiltonian in Eq. 8 possess the negative values, the diagonal on-site terms of the effective Hamiltonian of the circuit in Eq. 10 hold the term $t_{on}$ significantly greater than the difference $t_\Omega$ of the step variation. This mechanism ensures the eigenvalues of the Hamiltonian $\mathcal{H}(k)$ are positive.

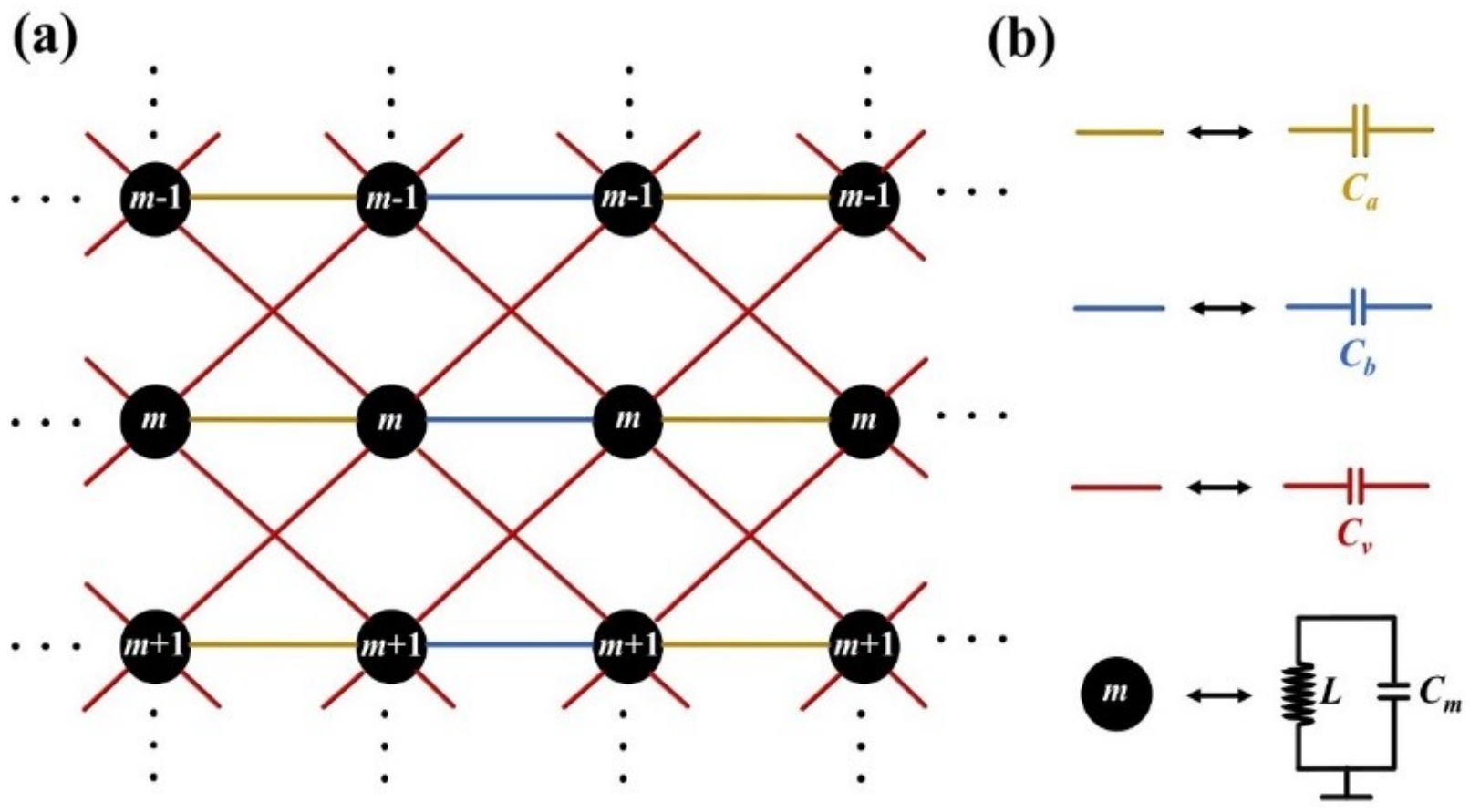

Figure 1. The schematic diagram of the Floquet circuit. **a**, the structure of the electric elements. In each row, the circuit is constructed as the SSH chain, with staggered coupling between the nearest-neighbor rows indicated by the orders $\{m \mid m \in [-M, M] \text{ and } m \in \mathbb{Z}\}$. **b**, the element list of the circuit. The components in each row include the intra- and inter-coupling $C_a$ and $C_b$, as well as the grounded inductor $L$ and capacitor $C_m$. The interlayer capacitor $C_v$ connects $a$ nodes to $b$ nodes in adjacent rows along columns, and vice versa. The lattice in each row contains two nodes connected by the intra-coupling capacitor $C_a$, while the hopping term between lattices

is governed by the inter-coupling capacitor $C_b$.

## The π and zero modes in the circuit

The band structure of $\mathcal{H}(k)$ is depicted in Fig. 2a-2c. Bands of different orders are distinguished by background colors. When the DF capacitor $t_\Omega$ is greater than the static band $W = 2(t_a + t_b)$ of the SSH model in each row of the circuit and $t_a$ is greater than $t_b$, the band gaps between bands of different orders (referred to as inter-band gaps) are approximately $t_\Omega - W$, as shown in Fig. 2a. Correspondingly, the band gaps within bands of each order are named intra-band gaps. When $t_\Omega$ equals $W$, frequency degeneracies arise between the bands with orders $m$ = 0 and ±1 (yellow- and green-background bands), as shown in Fig. 2b. The intra-band gaps with $m$ = ±1 and ±2 (green- and blue-background bands) remain open due to the hopping terms $H_{\pm 1}$. If the APM capacitor $t_v$ is much less than the intralayer coupling capacitors $t_a$ and $t_b$, the frequency intervals between adjacent bands are approximately equal to $|t_\Omega - W|$, and all the inter-band gaps approach zero. Comparatively, when the APM capacitor increases relative to the intralayer-coupling capacitors by $t_v \to t_a(t_b)$, some inter-band gaps remain open (detailed analysis is provided in Section 1 of the Appendix). When $t_\Omega$ is lower than $W$, previously closed inter-band gaps reopen as shown in Fig. 2c. According to the above analysis, the process of inter-band gap opening, closing, and reopening with the gradual change of the DF capacitors reveals the existence of a phase transition at $t_\Omega \approx W$.

Next, considering the circuit with open boundaries and a lattice size of 10 in each row, the Hamiltonian in Eq. 10 is transformed into its real-space form. The spectra with different values of the DF capacitors are shown in Fig. 2d-2f. When $t_\Omega$ is greater than or equal to $W$, the resonant frequencies of the circuit with open boundaries are all located within the band, as illustrated in Fig. 2d-2e. When $t_\Omega$ is lower than $W$, some additional resonant frequencies emerge in the inter-band gaps, as in Fig. 2f. According to the analysis in Section 2 of the Appendix, the states within the interlayer bandgap present the edge-bulk correspondence and correspond to the π modes of the FTIs. Afterward, we consider the condition $t_a < t_b$. The bands of $\mathcal{H}(k)$ are shown as Fig. 2d-2f. The resonant frequencies of the circuit with open boundaries emerge within the intralayer gaps and are not affected by the relationship between $t_\Omega$ and $W$, as shown in Fig. 2j-2l. According to the analysis in Section 2 and 3 in Appendix, the states within intralayer gaps also exhibit the edge-bulk correspondence and correspond to the zero modes of the FTIs.

Here we consider the condition $t_\Omega < W$ and $t_a < t_b$, the spectrum of the circuit with open boundaries exhibits the resonant frequencies of π and zero modes within the bandgap, as shown is shown in Fig. 3a. We calculate the states of π modes (indicated by $\phi_\pi^{(m\leftrightarrow m+1)}$ with $m \in [-M, M-1]$ ) and zero modes (indicated by $\phi_0^{(m)}$ with $m \in [-M, M]$), and present the bulk-edge correspondence in Fig. 3b-3c. The rows of the circuit from 1st to 5th (from the top row to bottom row) correspond to the orders $m$ = -2 to 2. The states $\phi_\pi^{(m\leftrightarrow m+1)}$ of π modes distribute the edges of the two rows with the order $m$ and $m + 1$. Correspondingly, the states $\phi_0^{(m)}$ of zero modes distribute the edge of the row with the order $m$.

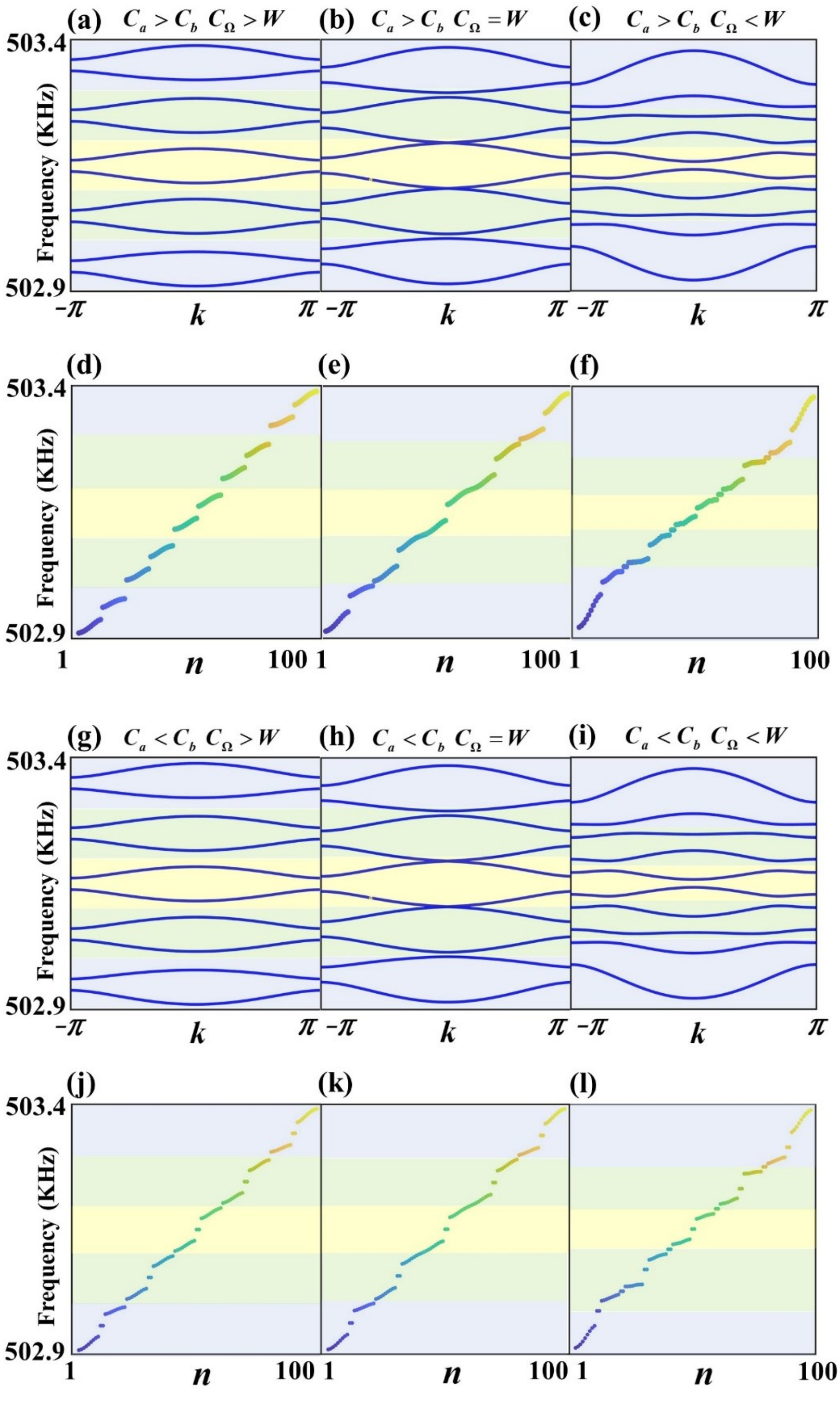

Figure 2. The spectra of the circuit. The values of the electric elements are $L = 100$ nH, $C_0 = 1$ μF and $C_v = 60$ pF. **a-c**, the band of the circuit with $C_a > C_b$. The DF capacitors are $C_\Omega = 440$ pF, 400 pF and 260 pF, while the intra-coupling capacitors are $C_a = 150$ pF and $C_b = 50$ pF. **d-f**, the resonant frequencies of the circuit with open boundaries. The DF capacitors are $C_\Omega = 440$ pF, 400 pF and 260 pF. **g-i**, the band of the circuit with $C_a < C_b$. the intra-coupling capacitors are $C_a = 50$ pF and $C_b = 150$ pF. **j-l**, the resonant frequencies of the circuit with open boundaries with $C_a = 50$ pF and $C_b = 150$ pF. The band and resonant frequencies of different Fourier orders are indicated by domains of different colors (blue, green, yellow), corresponding to the orders *m* = ±2, ±1, and 0. Here *n* represents the index of the resonant frequencies.

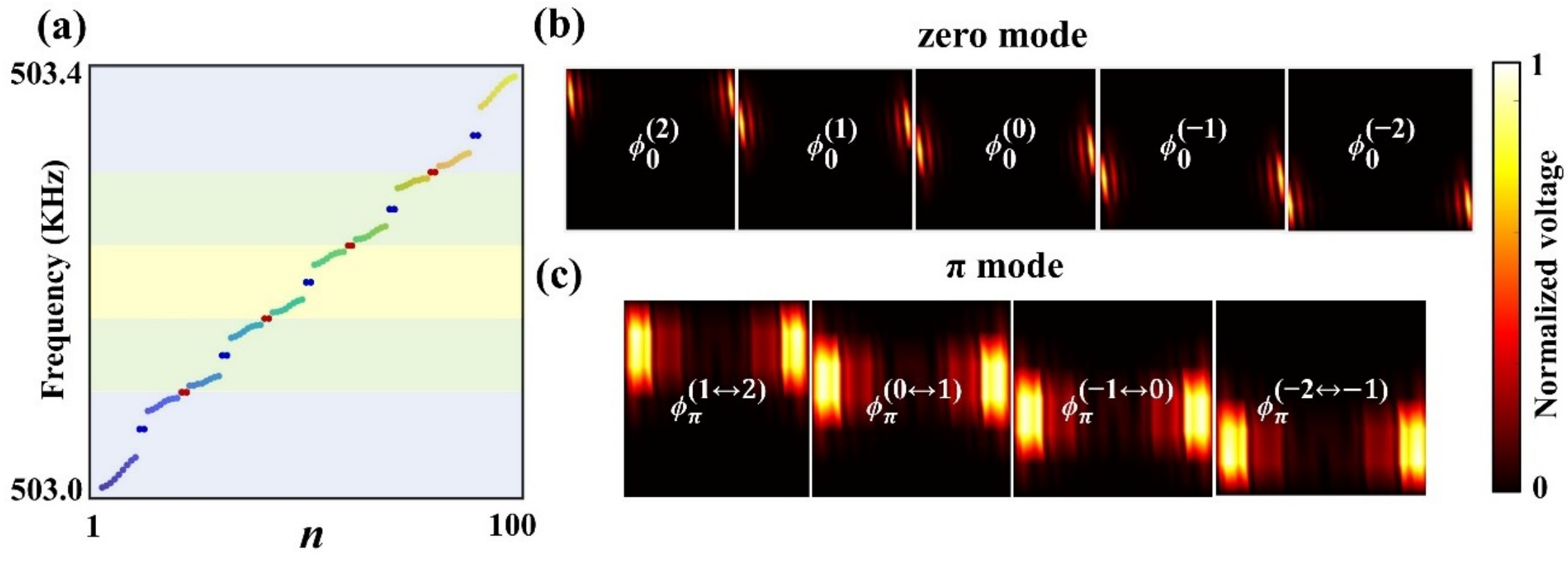

Figure 3. The spectrum and the spatial distribution of the zero and π mode. **a**, the spectrum of the circuit with $C_\Omega < W$ and $C_a < C_b$. The red and blue points represent the resonant frequencies of π modes and zero modes in the inter-band and intra-band gaps. **b-c**, the energy distribution of the zero- and π-modes states. The labels $\phi_0^{(m)}$ and $\phi_\pi^{(m\leftrightarrow m')}$ indicate states of zero modes at the intra-band gap of order *m* and π modes at the inter-band gap between the orders *m* and *m'*. The intra-layer coupling capacitors are $C_a = 50$ pF, $C_b = 150$ pF.

## The Subharmonic oscillations in the circuit

According to above analysis, the state on the edge node of the *m*-th row is approximately the superposition of the zero and π modes as

$$\left|\psi\right\rangle = \gamma_0 e^{-i\varepsilon_0^{(m)}t}\left|\phi_0^{(m)}\right\rangle + \alpha_\pi e^{-i\varepsilon_\pi^{(m-1\leftrightarrow m)}t}\left|\phi_\pi^{(m-1\leftrightarrow m)}\right\rangle + \beta_\pi e^{-i\varepsilon_\pi^{(m\leftrightarrow m+1)}t}\left|\phi_\pi^{(m\leftrightarrow m+1)}\right\rangle, \quad (11)$$

where $\gamma_0$, $\alpha_\pi$ and $\beta_\pi$ are the coefficients of $\left|\phi_0^{(m)}\right\rangle$, $\left|\phi_\pi^{(m-1\leftrightarrow m)}\right\rangle$ and $\left|\phi_\pi^{(m\leftrightarrow m+1)}\right\rangle$. The parameters $\varepsilon_{0/\pi}^{(m)/(m\leftrightarrow m')}$ are the frequencies of the zero and π modes. When the order is *m* = -*M* or *M*, the coefficients $\alpha_\pi$ or $\beta_\pi$ is zero due to the edge sites of the two rows possessing only one topological π mode $\left|\phi_\pi^{(M-1\leftrightarrow M)}\right\rangle$ or $\left|\phi_\pi^{(-M\leftrightarrow -M+1)}\right\rangle$.

The intensity of the states can be calculated as

$$\begin{aligned}\Psi &= \gamma_0^2\left\langle\phi_0^{(m)}\middle|\phi_0^{(m)}\right\rangle+\alpha_\pi^2\left\langle\phi_\pi^{(m-1\leftrightarrow m)}\middle|\phi_\pi^{(m-1\leftrightarrow m)}\right\rangle+\beta_\pi^2\left\langle\phi_\pi^{(m\leftrightarrow m+1)}\middle|\phi_\pi^{(m\leftrightarrow m+1)}\right\rangle\\ &+[\gamma_0\alpha_\pi e^{-i\left(\varepsilon_\pi^{(m-1\leftrightarrow m)}-\varepsilon_0^{(m)}\right)t}\left\langle\phi_0^{(m)}\middle|\phi_\pi^{(m-1\leftrightarrow m)}\right\rangle+\gamma_0\beta_\pi e^{-i\left(\varepsilon_\pi^{(m\leftrightarrow m+1)}-\varepsilon_0^{(m)}\right)t}\left\langle\phi_0^{(m)}\middle|\phi_\pi^{(m\leftrightarrow m+1)}\right\rangle\\ &+\alpha_\pi\beta_\pi e^{-i\left(\varepsilon_\pi^{(m\leftrightarrow m+1)}-\varepsilon_\pi^{(m-1\leftrightarrow m)}\right)t}\left\langle\phi_\pi^{(m-1\leftrightarrow m)}\middle|\phi_\pi^{(m\leftrightarrow m+1)}\right\rangle+\text{H.c.}].\end{aligned} \tag{12}$$

The frequencies of $\Psi$ in Eq. 12 are $\left|\varepsilon_\pi^{(m-1\leftrightarrow m)}-\varepsilon_0^{(m)}\right|\approx\left|\varepsilon_\pi^{(m\leftrightarrow m+1)}-\varepsilon_0^{(m)}\right|\approx\frac{t_\Omega}{2}$ and $\left|\varepsilon_\pi^{(m-1\leftrightarrow m)}-\varepsilon_\pi^{(m\leftrightarrow m+1)}\right|\approx t_\Omega$. In contrast to the DF capacitor $t_\Omega$, the signals on the edge nodes possess the resonances with the frequency $\frac{t_\Omega}{2}$. When only the π modes exist on the edge nodes, the intensity on the edge nodes only has the frequency $\left|\varepsilon_\pi^{(m-1\leftrightarrow m)}-\varepsilon_\pi^{(m\leftrightarrow m+1)}\right|\approx t_\Omega$. The theoretical verification is in Section 4 in Appendix.

In circuit system, the frequency of the periodically driven signal can be obtained from the difference of the resonant frequencies of the grounded-*LC* resonators on the adjacent rows as $\omega_\Omega=\omega_{0_m}-\omega_{0_{m-1}}=\frac{1}{\sqrt{L\left(C_0+mC_\Omega\right)}}-\frac{1}{\sqrt{L\left(C_0+(m+1)C_\Omega\right)}}$. Using the relation $C_\Omega << C_0$, the frequency of the periodically driven signal can be expressed as:

$$\begin{aligned}\omega_\Omega &= \omega_{0_m}-\omega_{0_{m+1}}=\frac{1}{\sqrt{L\left(C_0+mC_\Omega\right)}}-\frac{1}{\sqrt{L\left(C_0+(m+1)C_\Omega\right)}}\\ &=\frac{1}{\sqrt{LC_0}}\left(\frac{\sqrt{1+\frac{(m+1)C_\Omega}{C_0}}-\sqrt{1+\frac{mC_\Omega}{C_0}}}{\sqrt{1+\frac{mC_\Omega}{C_0}}\sqrt{1+\frac{(m+1)C_\Omega}{C_0}}}\right).\end{aligned} \tag{13}$$

We use Taylor series to simplify the numerator in parentheses, while the denominator is nearly equal to 1:

$$\sqrt{1+\frac{mC_\Omega}{C_0}}=1+\frac{1}{2}\frac{mC_\Omega}{C_0}+o\left(\frac{mC_\Omega}{C_0}\right),$$
$$\sqrt{1+\frac{mC_\Omega}{C_0}}=1+\frac{1}{2}\frac{(m+1)C_\Omega}{C_0}+o\left(\frac{(m+1)C_\Omega}{C_0}\right) \tag{14}$$

where $o(x)$ represents the infinitesimal when $x$ is close to zero. We substitute Eq. 14 into Eq. 13:

$$\omega_\Omega=\omega_{0_m}-\omega_{0_{m+1}}\approx\frac{1}{\sqrt{LC_0}}\left(\frac{C_\Omega}{2C_0}\right) \tag{15}$$

which states that the frequencies of the periodically driven signals on different rows are approximately equal under the condition $C_\Omega << C_0$.

The circuit is simulated using PathWave software. The source of inspiration is set to a constant current with an amplitude of 1 A in the frequency domain. This current is applied to site 1 (the edge node in the first row) of the circuit, and the results are shown in Fig. 4a, b. The result in Fig. 4a shows that a voltage peak is observed at 503.080 KHz which corresponds to the resonant frequency of the zero mode within the intralayer band gap, ranging from 503.055 to 503.096 KHz. The large difference between the edge and bulk voltages is due to the neglect of circuit element losses in this analysis. Thus, the expression of the voltage amplitude is presented using logarithmic formulation. The voltage distribution of the zero mode is shown in Fig. 4c. Analogously, the peak at 503.113 kHz, located within the inter-layer band gap of 503.108 – 503.119 kHz in Fig. 4b, corresponds to the frequency of the π mode, and the voltage distribution is shown in Fig. 4d. The voltages of zero and π mode exhibit the topological boundary-bulk correspondence. Then, a step signal is applied at site 1, where the current is 0 *A* for times less than 1 μs and 1 *A* for times greater than 1 μs. The response voltage on site 1 in time domain is shown as Fig. 4e. The results show that the voltage exhibits resonance at 503 KHz and fluctuation at 33 Hz. The signal at site 1 is Fourier transformed, and the result is shown in Fig. 4f. From the results of the circuit simulation in Fig. 4e and 4f, the period of the subharmonic oscillation is calculated as 32.87 ms, which is approximately double the period of the equivalent periodic driving signal. The simulation results show that deeply subharmonic oscillations emerge in the presence of both zero and π modes.

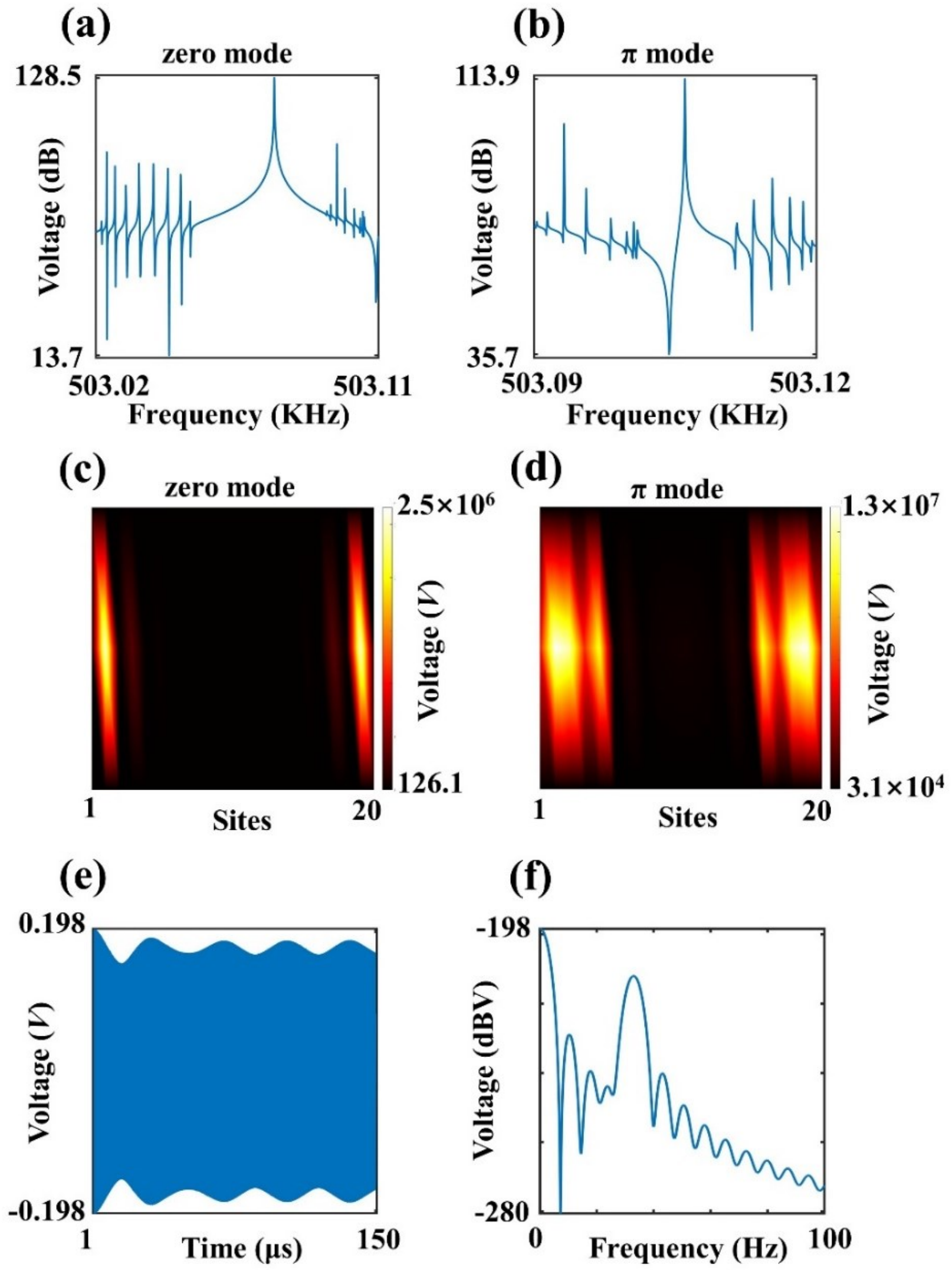

Figure 4. The simulation results of the circuit under the coexistence of zero and π modes. **a**-**b**, the voltages in frequency domain. **c**-**d**, the voltages of the zero and π modes. **e**, the voltages in time domain at site 1. **f**, the Fourier transformation of the voltage at site 1.

## Conclusion

In this paper, we propose the circuit to realize the Floquet Hamiltonian by constructing the frequency-synthetic dimension. The Floquet π mode and deeply subharmonic oscillation are investigated in the circuit system. Additionally, the band of the circuit with large time-modulation amplitude is studied. In the future, leveraging the flexibility of the circuit, we can construct non-Hermitian or non-linear Floquet topological systems based on the frequency-synthetic dimension.

**Acknowledgements**

B.L. was sponsored by the National Natural Science Foundation of China under Grant No. 61901133; Basic Research Business Expenses (3072024XX2504) J.S. was sponsored by National Natural Science Foundation of China (NSFC) (62275061, 62175049, U22A2014); Natural Science Foundation of Heilongjiang Province in China (ZD2020F002); 111 project to the Harbin Engineering University (B13015); Fundamental Research Funds for the Central Universities (3072022TS2509). F. M. was sponsored by National Natural Science Foundation of China (NSFC) under Grant No. U22A2014.

**Data availability**

The data that support the findings of this study are available from the corresponding author upon reasonable request.

**AUTHOR CONTRIBUTIONS**

B. L. conceived the idea. B. L., S. X., Y. T., T. L., H. M., Z. S. and S. W. carried out the calculations of the Floquet bands. B. L., F. M., Z. Z. and J. S. wrote the article. H. T. provides the software PathWave for simulating the circuit. Besides, we are extremely grateful to Shunyu Yao at SITP for his help in topological-invariant discussion and Qingqing Cheng at University of Shanghai for Science and Technology for his help in the time-evolution operator calculation.

**APPENDIX:**

**Section 1. The band of the Hamiltonian with different APM capacitors**

Based on Kirchhoff's law, we can derive the dynamic equations for the circuit shown in Fig. 1 in momentum space as

$$\begin{cases} I_{m1} = C_a \frac{d(V_{m1}-V_{m2})}{dt} + C_b \frac{d(V_{m1}-V_{m2}e^{-ik})}{dt} + C_m \frac{dV_{m1}}{dt} + \frac{1}{L_0}\int V_{m1}dt + C_v \frac{d(V_{m1}-V_{m-12})}{dt} + C_v \frac{d(V_{m1}-V_{m-12}e^{-ik})}{dt} + C_v \frac{d(V_{m1}-V_{m+12})}{dt} + C_v \frac{d(V_{m1}-V_{m+12}e^{-ik})}{dt} \\ I_{m2} = C_a \frac{d(V_{m2}-V_{m1})}{dt} + C_b \frac{d(V_{m2}-V_{m1}e^{ik})}{dt} + C_a \frac{dV_{m2}}{dt} + \frac{1}{L_0}\int V_{m2}dt + C_v \frac{d(V_{m2}-V_{m-11})}{dt} + C_v \frac{d(V_{m2}-V_{m-11}e^{ik})}{dt} + C_v \frac{d(V_{m2}-V_{m+11})}{dt} + C_v \frac{d(V_{m2}-V_{m+11}e^{ik})}{dt} \end{cases}.$$

(A1)

Here we set the total currents at the nodes to zero and differentiate the voltages of the nodes within one-unit cell with respect to time *t*. The equations are expressed as

$$\begin{cases} -\frac{V_{m1}}{C_0L_0} = (1+t_a+t_b+mt_\Omega)\frac{d^2V_{m1}}{dt^2} + (-t_a-t_be^{-ik})\frac{d^2V_{m2}}{dt^2} + t_v(1+e^{-ik})\frac{d^2V_{m-12}}{dt^2} + t_v(1+e^{-ik})\frac{d^2V_{m+12}}{dt^2} \\ -\frac{V_{m2}}{C_0L_0} = (1+t_a+t_b+mt_\Omega)\frac{d^2V_{m2}}{dt^2} + (-t_a-t_be^{ik})\frac{d^2V_{m1}}{dt^2} + t_v(1+e^{ik})\frac{d^2V_{m-11}}{dt^2} + t_v(1+e^{ik})\frac{d^2V_{m+11}}{dt^2} \end{cases}.$$

(A2)

We substitute $\omega_0^2 = \frac{1}{C_0L_0}$ and Fourier transform $\frac{d^n f(t)}{dt^n} = (j\omega)^n F(\omega)$ into the equations and obtain the dynamic equations of the circuit as

$$\begin{cases} \frac{\omega_0^2}{\omega^2}V_{m1} = (1+t_a+t_b+mt_\Omega)V_{m1} + (-t_a-t_be^{-ik})V_{m2} + t_v(1+e^{-ik})V_{m-12} + t_v(1+e^{-ik})V_{m+12} \\ \frac{\omega_0^2}{\omega^2}V_{m2} = (1+t_a+t_b+mt_\Omega)V_{m2} + (-t_a-t_be^{ik})V_{m1} + t_v(1+e^{ik})V_{m-11} + t_v(1+e^{ik})V_{m+11} \end{cases}. \quad (A3)$$

This equation can be simplified to the form $\frac{\omega_0^2}{\omega^2}V = HV$, and the Hamiltonian of the circuit takes the form given in Eq. 6.

The spectrum of the Hamiltonian in Eq. 10 is calculated under different conditions of the APM capacitor $t_v$, as shown in Fig. 5. When the APM capacitors are set to 10 pF, which is much smaller than the intralayer-coupling capacitors $t_a$/$t_b$, the hopping terms $H_{\pm 1}$ can be regard as very small perturbation, and all the inter-band gaps are almost closed, as shown in Fig. 5a. When the APM capacitor increases to $t_v = t_a$, the inter-band gaps between the bands of the orders $m = \pm 1$ and $m = \pm 2$ (between the green and blue domains) remain open, as shown in Fig. 5b. When the APM capacitor increases to 100 pF, all the inter-band gaps open, as shown in Fig. 5c. If we aim to close the inter-band gaps, we need to increase the value of the DF capacitors. When the APM capacitor increases to 300 pF, which is higher than the intra-layer capacitors, the bands of adjacent orders have some coincident eigenvalues with different *k*, resulting in the circuit having intra-layer gaps only in the band of order $m = 0$, as shown in Fig. 5d. When the APM capacitor increases to 1 nF, the intralayer bands of each order gradually getting closer together, as shown in Fig. 5e. When the APM capacitor increases to 1 μF, all the intralayer bands of each order complete overlap, and the bands of the order $m = 0$ exhibit a flat-band form, as shown in Fig. 5f. Here, the bands with different orders are represented with different background colors.

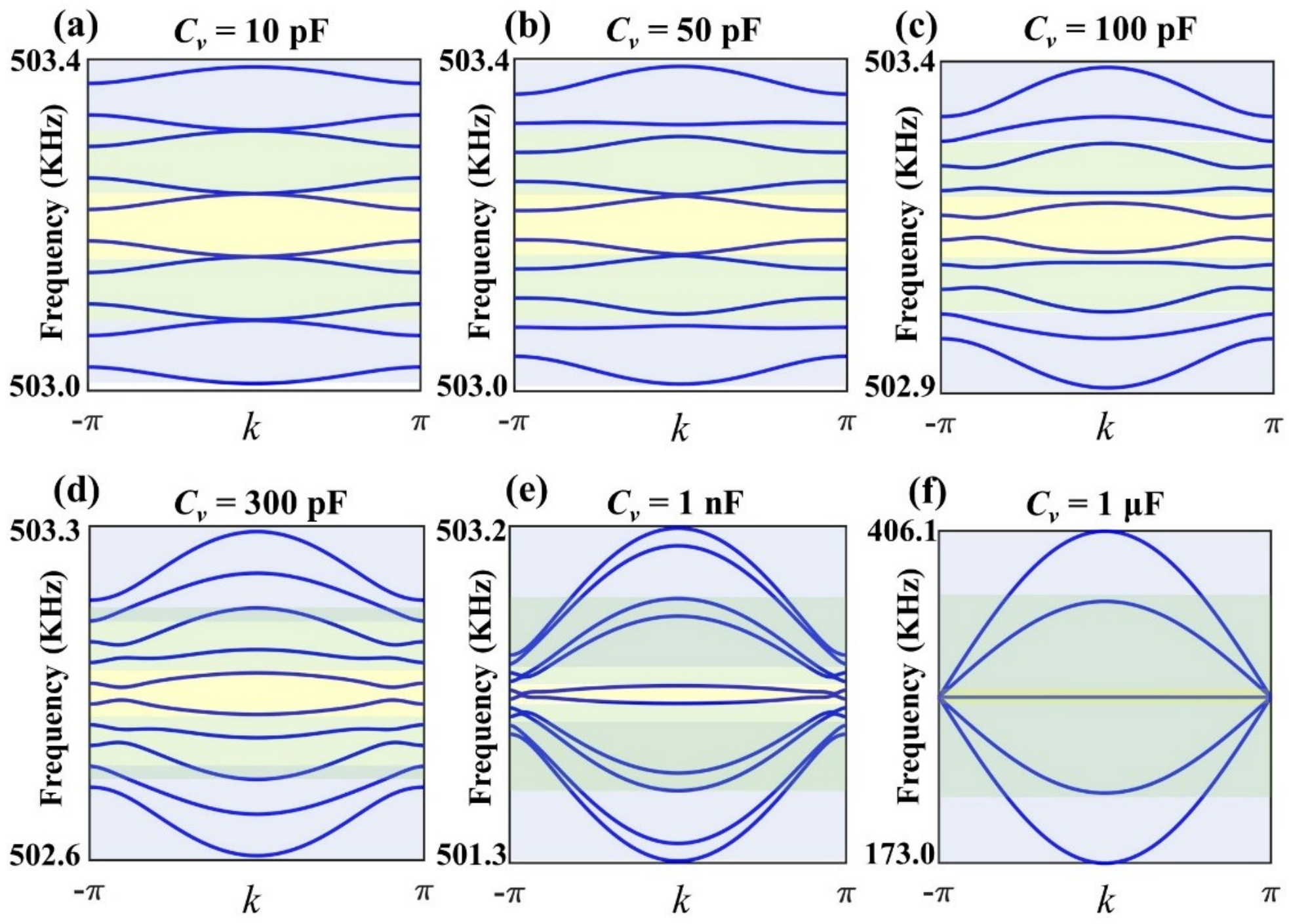

Figure 5. The bands of the Hamiltonian with different APM capacitors as $C_v$ = 10 pF, 50 pF, 100 pF, 300 pF, 1 nF and 1 μF. The values of other electric elements are set as $L = 100$ nH, $C_0 = 1$ μF, $C_a = 100$ pF, $C_b = 30$ pF and $C_\Omega = 300$ pF. The domains with different colors indicate different orders (he blue/green/yellow domains correspond to the orders $m = \pm 2/\pm 1/0$).

**Section 2. The time-evolution performance of the Hamiltonian**

In this section, we research the time evolution of the Hamiltonian in Eq. 1 and illustrate that the eigenvalues in the interlayer gaps correspond to the topologically protected π mode [13]. The time-evolution operator from $t_0$ to $t$ is expressed as

$$U(t,t_0) = \hat{T}\exp\left(-i\int_{t_0}^{t} H(t')\,dt'\right), \tag{A4}$$

where $\hat{T}$ denotes time ordering. The time-evolution operator is the solution of the dynamic equation given by

$$\left(H(t)-i\partial_t\right)U(t,t_0)=0. \tag{A5}$$

Here we define the initial time $t_0$ = 0, and express time-evolution operator as $U(t)=U(t,0)=\hat{T}\exp\left(-i\int_0^t H(t')dt'\right)$. The effective Hamiltonian is defined as

$$H_{eff}\equiv\frac{i}{T}\ln U(T). \tag{A6}$$

The eigenvalues of $H_{eff}$ localized in the Floquet Brillouin zone [-π/$T$, π/$T$] are indicated as $\varepsilon_{eff}$, which is shown in Fig. 6a.

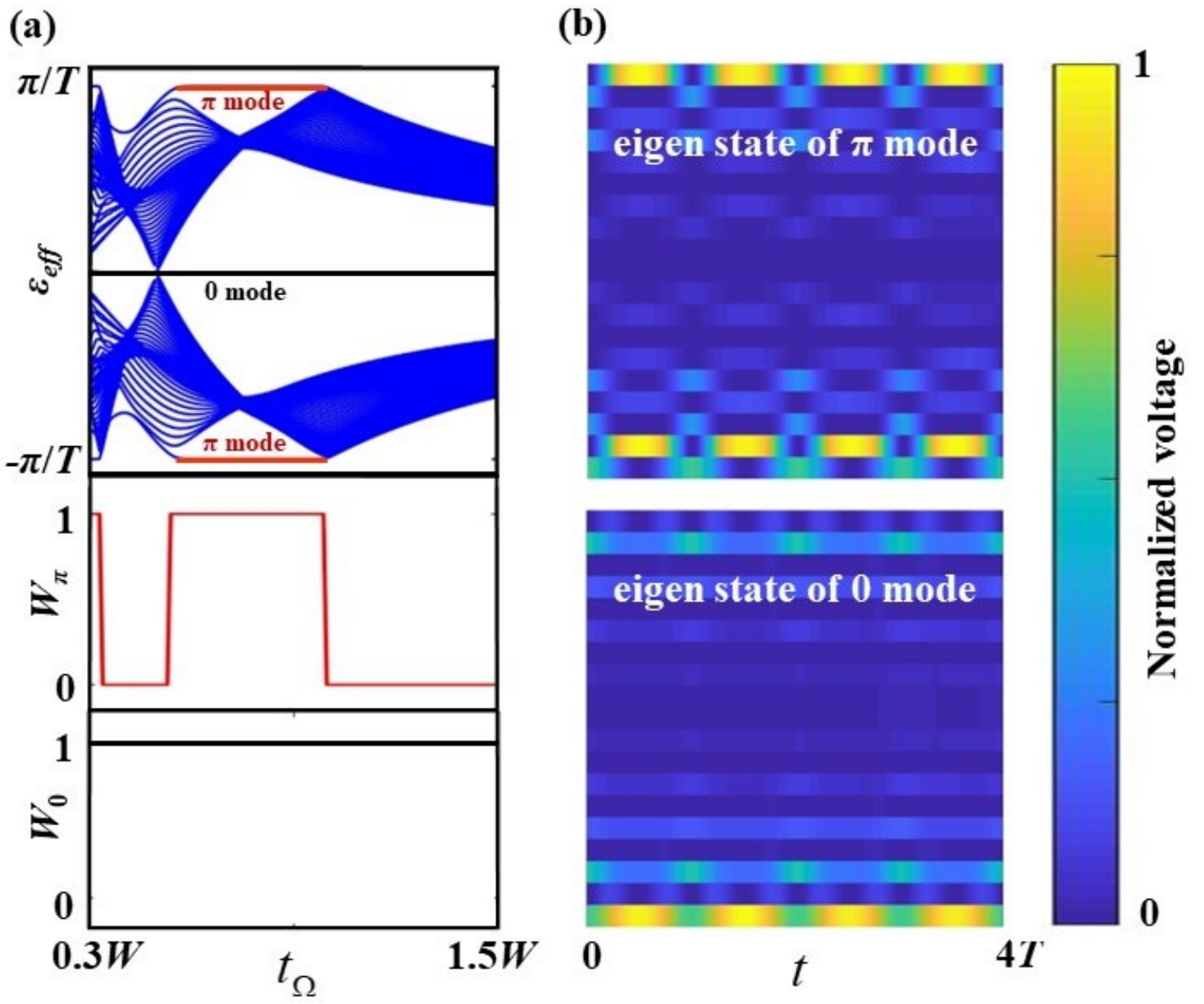

Figure 6. The spectra and states of the effective Hamiltonian. **a**, the spectra of the effective Hamiltonian in Floquet Brillouin zone [-π/$T$, π/$T$]. The red and black lines indicate the quasienergies of the π and zero modes. The corresponding topological invariants $W_{\pi/0}$ of the π and zero modes are also calculated. **b**, the time evolution of the π and zero modes of the circuit with 40 lattice sites. The main energy of the two topological-protected states is concentrated at the boundaries. The values of the electric elements are set as $L=100$ nH, $C_0=1$ μF, $C_a=30$ pF, $C_b=100$ pF and $C_v=50$ pF, and the DF capacitor $t_\Omega$ is set in the range of [0.3$W$, 1.5$W$].

Here we only discuss the topological performance under the condition of $t_\Omega\geq 0.5W$. Figure 6a shows that the π mode is located at $t_\Omega\in[0.55W, 0.99W]$ and the zero mode always exists throughout the entire $t_\Omega$ range. Comparing with the static bandwidth $W$, the error of the $\pi$-mode topological-phase transition point, $0.01W$, arises from the effect of the hopping $H_{\pm 1}$ in the second diagonal terms of the Floquet Hamiltonian. Moreover, we investigate the topological properties of the zero and $\pi$ modes by solving the chiral gap invariants of the periodic-driven model. We define a dimensionless quasienergy $\varepsilon' = \varepsilon_{eff} T$, which is used to define the branch out of the effective Hamiltonian as

$$H_{eff}^{\varepsilon'} = \frac{i}{T} \ln_{-\varepsilon'} U(T), \tag{A7}$$

where the complex logarithm $\ln_{\alpha} e^{i\phi}$ with the branch out satisfies the relation $\ln_{\alpha} e^{i\phi} = i\phi$ for $\alpha - 2\pi < \phi < \alpha$. If we regard the time evolution of the Hamiltonian $H(t)$ as the changes in the phase $\phi(t)$, the logarithm in Eq. A7 indicates that the information of the time evolution is compressed within the range $\varepsilon' - 2\pi < \phi < \varepsilon'$. Obviously, the Hamiltonian $H_{eff}$ in Eq. A3 corresponds to the effective Hamiltonian in Eq. A7 with $\varepsilon' = \pi$. In the follow content in this section, we use the form $H_{eff}^{\varepsilon'}$ for expressing the effective Hamiltonian with the branch out $\varepsilon'$. From the expression Eq. A7, the effective Hamiltonian $H_{eff}^{\varepsilon'}$ cannot be used to solve the topological invariants because it only captures the stroboscopic evolution at integer multiples of $T$ and lacks information within the period. Although the operator $U(t)$ can contain the evolution within the period, it is not periodic itself as $U(t) \neq U(t+T)$. Here we define a periodized evolution operator given by

$$V_{\varepsilon'}(t) = U(t) \exp\left(-iH_{eff}^{\varepsilon'} t\right), \tag{A8}$$

which satisfies the discrete $T$-translational symmetry $V_{\varepsilon'}(t) = V_{\varepsilon'}(t+T)$ and contains the short-timescale information. Using these features of the operator $V_{\varepsilon'}(t)$, we can construct the relation between the Floquet Hamiltonian $H_F$ and the effective Hamiltonian $H_{eff}^{\varepsilon'}$ as

$$V_{\varepsilon'}^{-1}(t)H_F V_{\varepsilon'}(t)=V_{\varepsilon'}^{-1}(t)\left(H(t)-\partial_t^2\right)V_{\varepsilon'}(t)H_{eff}^{\varepsilon'}. \tag{A9}$$

The periodized evolution operator $V_{\varepsilon'}(t)$ is considered to be the unitary rotation matrix. This transformation relation suggests that the stroboscopic evolution of the time-periodic-driven system can be analyzed using either the Floquet Hamiltonian or the effective Hamiltonian. For static systems, the topological invariant typically arises from the summation over all bands below the Fermi level. However, for the time-periodic driven system, the Fourier replicas of the Floquet bands are infinite, and a strict definition of the Fermi level does not exist. Thus, in this article, we utilize the winding number $W_{\varepsilon'}$, defined in terms of chiral gaps $\varepsilon_{eff}=0$ or π as the topological invariant. According to the chiral symmetry $\Gamma H(k,t)\Gamma^{-1}=-H(k,-t)$, the periodized evolution operator can be expressed as

$$\Gamma V_{\varepsilon'}(t,k)\Gamma^{-1}=-V_{-\varepsilon'}(-t,k)\exp(2\pi it/T) \tag{A10}$$

The zero mode corresponds to the branch out $\varepsilon'=0$. The chiral constraint is expressed as

$$\Gamma V_0\left(\frac{T}{2},k\right)\Gamma^{-1}=-V_0\left(\frac{T}{2},k\right), \tag{A11}$$

where the operator has the antidiagonal form

$$V_0\left(\frac{T}{2},k\right)=\begin{pmatrix}0 & V_0^+\\ V_0^- & 0\end{pmatrix}. \tag{A12}$$

The winding number for zero mode is determined by

$$W_0=\frac{i}{2\pi}\int_{-\pi}^{\pi}tr\left[\left(V_0^+\right)^{-1}\partial k V_0^+\right]dk. \tag{A13}$$

For the π mode corresponding to $\varepsilon'=\pi$, the periodized evolution operator under chiral constraint is the form

$$\Gamma V_\pi\left(\frac{T}{2},k\right)\Gamma^{-1}=V_\pi\left(\frac{T}{2},k\right), \tag{A14}$$

where the operator is the diagonal form

$$V_\pi\left(\frac{T}{2},k\right)=\begin{pmatrix}V_\pi^+ & 0\\ 0 & V_\pi^-\end{pmatrix}. \tag{A15}$$

The winding number for π mode is calculated using

$$W_{\pi} = \frac{i}{2\pi}\int_{-\pi}^{\pi} tr\left[\left(V_{\pi}^{+}\right)^{-1}\partial k V_{\pi}^{+}\right]dk. \tag{A16}$$

The winding numbers $W_{\pi}$ and $W_0$ for different driven frequencies are shown in Fig. 7a. For the π mode, topological-protected states emerge under the condition of $t_{\Omega} \in [0.55W, 0.99W]$, corresponding to the Winding number $W_{\pi} = 1$. The topological changes indicated by Winding number further illustrate the phase transition in π gap of the spectrum of $H_{eff}$. Similar to the π mode, the topological characteristic of zero mode indicated by Winding number, which reads $W_0 = 1$, also corresponds to the nontrivial-topological features in zero gap.

The time evolution of the π and zero modes is depicted in Fig. 6b. The energy of the nontrivial modes is primarily concentrated at the edge nodes. The dynamic frequencies of π and zero modes are equal to the driven frequency. This energy-spatial distribution illustrates the bulk-boundary correspondence. In the subsequent content, subharmonic oscillation is observed at the edge nodes under the coexistence of π and zero modes.

**Section 3. The dynamics of the circuit with only zero modes**

Here, we investigate the topological edge states of the circuit featuring only zero modes. Then, we designate the intralayer capacitors as $t_a < t_b$ and the driven capacitors as $t_{\Omega} > W$. Under this condition, the system only possesses the zero modes. The spectra and voltage distribute of the zero modes are shown in Fig. 7a and Fig. 7b, respectively. The results show that the states of the zero modes exhibit edge-bulk correspondence in each row of the circuit.

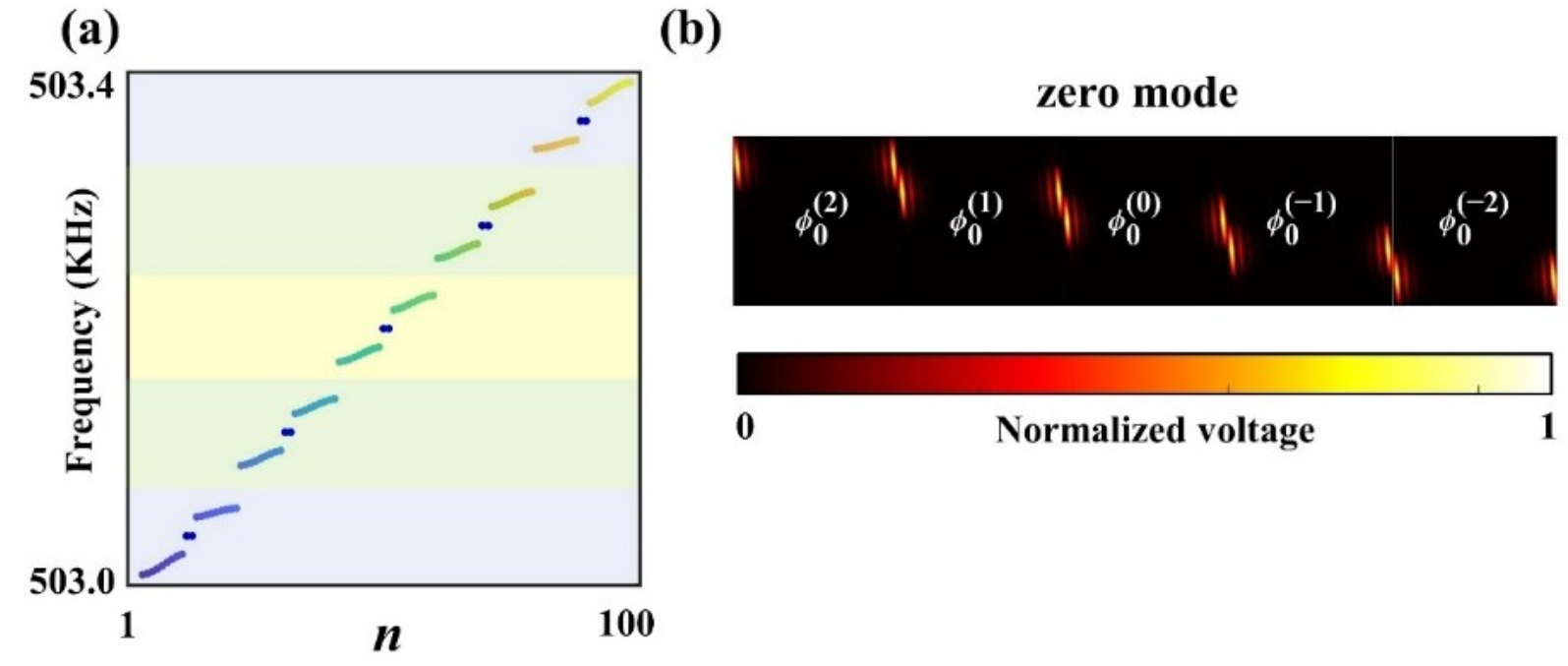

Figure 7. The spectrum in the presence of only the zero mode and the spatial distribution of the zero-mode states. **a**, the spectrum of the Floquet Hamiltonian with open boundary. The blue points within the intra-band gaps represent the eigenvalues of zero modes. **b**, the voltage distribution of the states of zero-modes. The labels $\phi_0^{(m)}$ indicate states of the zero mode within the intra-band gap of the *m*-th band.

The topological characteristics of the zero and π modes in circuit-SSH model are related to the values of the $t_{a/b}$, $t_\Omega$ and $W$. The topological invariants of the two modes are calculated using Eq. A13 and Eq. A16 in Appendix Section 2. When the topological characteristics of the zero and π modes are either nontrivial or trivial, the edge states of the circuit are depicted in Fig. 8. This figure illustrates the merging of topological edge-bulk correspondence with nontrivial topological performance.

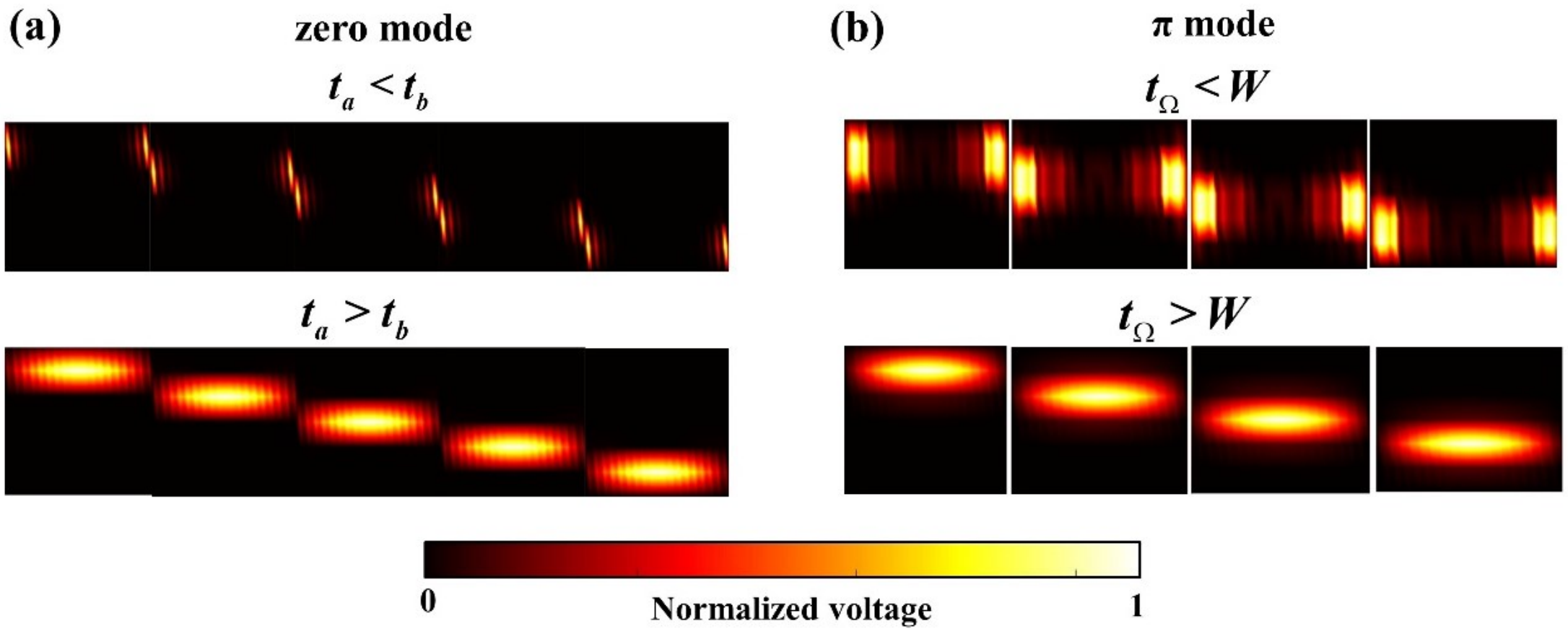

Figure 8. The spatial distribution of states with nontrivial and trivial characteristics for (a) zero modes and (b) π modes in the Floquet circuit. The lattice consists of 10 rows and 5 columns.

**Section 4. The response signals on the edge nodes**

We consider a time crystal with the driven frequency $t_\Omega$. Under the coexistence of the zero and π mode, the signals at site 1 (the edge node of the crystal) are calculated as the red lines in Fig. 9a. By contrast, the equivalent time-modulation signal with the driven frequency $t_\Omega$ and the signal at the site 2 (the bulk node of the crystal) are represented by the black-dash and blue lines. Comparing with the driving period $T = \frac{2\pi}{t_\Omega}$, the results illustrate that the double-periodic resonance emerges at the edge node. Furthermore, the Fourier transforms of the periodic-driven and the response signals at site 1 are shown as the black-dash and red lines in Fig. 9c, illustrating that the edge signal exhibits resonance at half the frequency of the periodic-driven signal. When the model possesses only the π mode at the edge nodes, the signals at sites 1 and 2 shown in Fig. 9b illustrate that the double-periodic resonance disappears at edge nodes. Additionally, the Fourier transforms of the response signals are shown in Fig. 9d. The signals at the edge nodes of other rows are also calculated in in the Section 4 in Appendix.

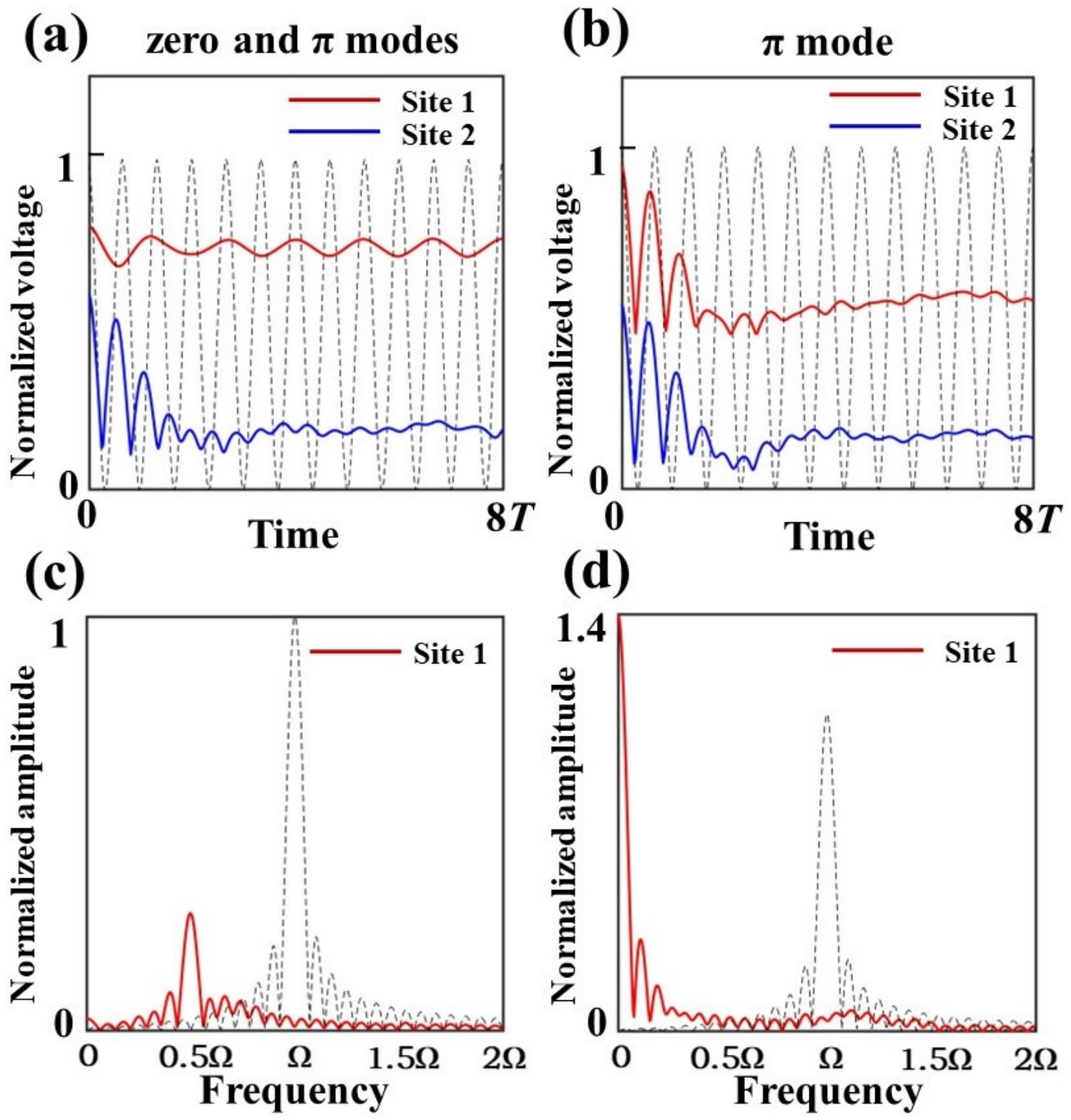

Figure 9. The periodically driven and response signals at the edge and bulk nodes. **a**-**b**,

the response signals at site 1 (the edge node) and site 2 (the bulk node) under the coexistence of zero and π modes or only the existence of π mode. **c-d** the Fourier transformation of the periodically driven and response signals under the coexistence of zero and π modes or only the existence of π mode.

The response signals at the edge nodes of different rows exhibit double-periodic resonance. Fig. 10a presents the double-periodic resonances at the edge nodes of the 1st to 5th rows under the condition of coexistence of zero and π modes. When the zero mode is nonexistent, the response signals at the edge nodes of different rows are presented in Fig. 10b.

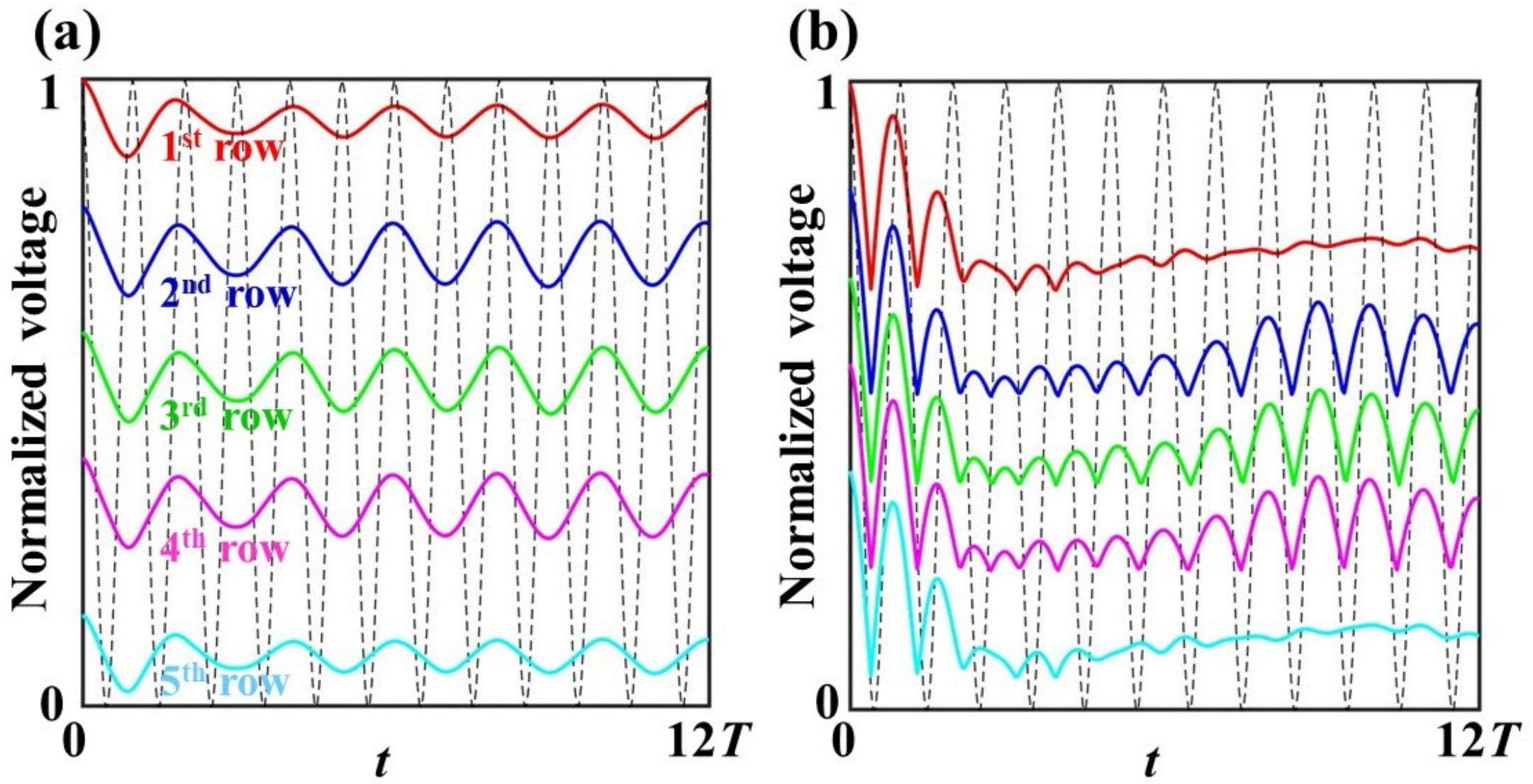

Figure 10. The response signals on the edge nodes of different rows. **a**, the response signals at the edge nodes under the condition of coexisting zero and π modes. The red, blue, green, purple, and cyan lines indicate the response signals at the edge nodes of the $1^{st}$, $2^{nd}$, $3^{rd}$, $4^{th}$, and $5^{th}$ rows, corresponding to the $m$ = 2/1/0/-1/-2 orders. **b**, The response signals on the edges under the condition of the only existence of π modes. The colors of the lines correspond to the same orders as in Fig. 10a. The dashed line indicates the signal with the driven frequency equal to the DF capacitor $t_\Omega$.

## Reference

[1] M. Z. Hasan and C. L. Kane, *Colloquium*: Topological insulators, Rev. Mod. Phys 82, 3045 (2010).

[2] J. E. Moore, The birth of topological insulators, Nature 464, 194 (2010).
[3] Y. Tokura, K. Yasuda and A. Tsukazaki, Magnetic topological insulators, Nat Rev Phys 1, 126 (2019).
[4] L. Fu, C. L. Kane and E. J. Mele, Topological Insulators in Three Dimensions, Phys. Rev. Lett 98, 106803 (2007).
[5] W. –L. Qi and S. –C. Zhang, Topological insulators and superconductors, Rev. Mod. Phys 83, 1057 (2011).
[6] M. C. Rechtsman, J. M. Zeuner, Y. Plotnik, Y. Lumer, D. Podolsky, F. Dreisow, S. Nolte, M. Segev and A. Szameit, Photonic Floquet topological insulators, Nature 496, 196 (2013).
[7] M. S. Rudner and N. H. Lindner, Band structure engineering and non-equilibrium dynamics in Floquet topological insulators, Nat Rev Phys 2, 229 (2020).
[8] R. Roy and F. Harper, Periodic table for Floquet topological insulators, Phys. Rev. B 96, 155118 (2017).
[9] S. K. Ivanov, Y. Zhang, Y. V. Kartashov and D. V. Skryabin, Floquet topological insulator laser, APL Photonics 4, 126101 (2019).
[10] S. Yao, Z. Yan and Z. Wang, Topological invariants of Floquet systems: General formulation, special properties, and Floquet topological defects, Phys. Rev. B 96, 195303 (2017).
[11] Q. Cheng, Y. Pan, H. Wang, C. Zhang, D. Yu, A. Gover, H. Zhang, T. Li, L. Zhou, and S. Zhu, Observation of Anomalous π Modes in Photonic Floquet Engineering, Phys. Rev. Lett. 122, 173901 (2019).
[12] S. Wu, W. Song, S. Gao, Y. Chen, S. Zhu and T. Li, Floquet π mode engineering in non-Hermitian waveguide lattices, Phys. Rev. Research 3, 023211 (2021).
[13] Y. Pan and B. Wang, Time-crystalline phases and period-doubling oscillations in one-dimensional Floquet topological insulators, Phys. Rev. Research 2, 043239 (2020).
[14] B. Wang, J. Quan, J. Han, X. Shen, H. Wu and Y. Pan, Observation of Photonic Topological Floquet Time Crystals, LASER PHOTONICS REV lpor. 202100469 (2022).
[15] L. Jiang, T. Kitagawa, J. Alicea, A. R. Akhmerov, D. Pekker, G. Refael, J. I. Cirac,

E. Demler, M. D. Lukin, and P. Zoller, Majorana Fermions in Equilibrium and in Driven Cold-Atom Quantum Wires, Phys. Rev. Lett. 106, 220402 (2011).

[16] M. Thakurathi, A. A. Patel, D. Sen, and A. Dutta, Floquet generation of Majorana end modes and topological invariants, Phys. Rev. B 88, 155133 (2013).

[17] A. Kundu and B. Seradjeh, Transport Signatures of Floquet Majorana Fermions in Driven Topological Superconductors, Phys. Rev. Lett. 111, 136402 (2013).

[18] P. Molignini, Wei Chen, and R. Chitra, Universal quantum criticality in static and Floquet-Majorana chains, Phys. Rev. B 98, 125129 (2018).

[19] X. Zhang and J. Gong, Non-Hermitian Floquet topological phases: Exceptional points, coalescent edge modes, and the skin effect, Phys. Rev. B 101, 045415 (2020).

[20] H. Dehghani, T. Oka and A. Mitra, Out-of-equilibrium electrons and the Hall conductance of a Floquet topological insulator, Phys. Rev. B 91, 155422 (2015).

[21] K. W. Kim, D. Bagrets, T. Micklitz and A. Altland, Quantum Hall criticality in Floquet topological insulators, Phys. Rev. B 101, 165401 (2020).

[22] H. Wang, L. Zhou, and Y. D. Chong, Floquet Weyl phases in a three-dimensional network model, Phys. Rev. B 93, 144114 (2016).

[23] L. Bucciantini, S. Roy, S. Kitamura and T. Oka, Emergent Weyl nodes and Fermi arcs in a Floquet Weyl semimetal, Phys. Rev. B 96, 041126(R) (2017).

[24] H. Hübener, M. A. Sentef, U. D. Giovannini, A. F. Kemper and A. Rubio, Creating stable Floquet–Weyl semimetals by laser-driving of 3D Dirac materials, Nat. Commun 8, 13940 (2017).

[25] Z. Yan and Z. Wang, Floquet multi-Weyl points in crossing-nodal-line semimetals, Phys. Rev. B 96, 041206(R) (2017).

[26] R. Chen, D.-H. Xu, and B. Zhou, Floquet topological insulator phase in a Weyl semimetal thin film with disorder, Phys. Rev. B 98, 235159 (2018).

[27] M. Umer, R. W. Bomantara, and J. Gong, Dynamical characterization of Weyl nodes in Floquet Weyl semimetal phases, Phys. Rev. B 103, 094309 (2021).

[28] S. Franca, J. van den Brink and I. C. Fulga, An anomalous higher-order topological insulator, Phys. Rev. B 98, 201114(R) (2018).

[29] M. R. -Vega, A. Kumar and B. Seradjeh, Higher-order Floquet topological phases

with corner and bulk bound states, Phys. Rev. B 100, 085138 (2019).
[30] T. Nag, V. Juričić, and B. Roy, Out of equilibrium higher-order topological insulator: Floquet engineering and quench dynamics, Phys. Rev. Research 1, 032045(R) (2019).
[31] Y. Peng, Floquet higher-order topological insulators and superconductors with space-time symmetries, Phys. Rev. Research 2, 013124 (2020).
[32] B. Huang and W. V. Liu, Floquet Higher-Order Topological Insulators with Anomalous Dynamical Polarization, Phys. Rev. Lett. 124, 216601 (2020).
[33] H. Hu, B. Huang, E. Zhao, and W. V. Liu, Dynamical Singularities of Floquet Higher-Order Topological Insulators, Phys. Rev. Lett. 124, 057001 (2020).
[34] D. Vu, R. –X. Zhang, Z. –C. Yang and S. D. Sarma, Superconductors with anomalous Floquet higher-order topology, Phys. Rev. B 104, L140502 (2021).
[35] T. Nag, V. Juričić and Bitan Roy, Hierarchy of higher-order Floquet topological phases in three dimensions, Phys. Rev. B 103, 115308 (2021).
[36] W. Zhu, Y. D. Chong and J. Gong, Floquet higher-order topological insulator in a periodically driven bipartite lattice, Phys. Rev. B 103, L041402 (2021).
[37] W. Zhu, H. Xue, J. Gong, Y. Chong and B. Zhang, Time-periodic corner states from Floquet higher-order topology, Nat. Commun 13, 11 (2022).
[38] M. D. Reichl and E. J. Mueller, Floquet edge states with ultracold atoms, Phys. Rev. A 89, 063628 (2014).
[39] Karen Wintersperger, Christoph Braun, F. Nur Ünal, André Eckardt, Marco Di Liberto, Nathan Goldman, Immanuel Bloch & Monika Aidelsburger Realization of an anomalous Floquet topological system with ultracold atoms. Nat. Phys. 16, 1058 (2020).
[40] P. Wang, Q. –F. Sun and X. C. Xie, Transport properties of Floquet topological superconductors at the transition from the topological phase to the Anderson localized phase, Phys. Rev. B 90, 155407 (2014).
[41] Y. Peng and G, Refael, Time-quasiperiodic topological superconductors with Majorana multiplexing, Phys. Rev. B 98, 220509(R) (2018).
[42] T. Čadež, R. Mondaini and P. D. Sacramento, Edge and bulk localization of Floquet topological superconductors, Phys. Rev. B 99, 014301 (2019).

[43] T. Čadež, R. Mondaini and P. D. Sacramento, Edge and bulk localization of Floquet topological superconductors, Phys. Rev. B 99, 014301 (2019).

[44] K. Plekhanov, M. Thakurathi, D. Loss and J. Klinovaja, Floquet second-order topological superconductor driven via ferromagnetic resonance, Phys. Rev. Research 1, 032013(R) (2019).

[45] L. Zhou, Non-Hermitian Floquet topological superconductors with multiple Majorana edge modes, Phys. Rev. B 101, 014306 (2020).

[46] A. K. Ghosh, T. Nag, and A. Saha, Floquet generation of a second-order topological superconductor, Phys. Rev. B 103, 045424 (2021).

[47] A. K. Ghosh, T. Nag, and A. Saha, Floquet second order topological superconductor based on unconventional pairing, Phys. Rev. B 103, 085413 (2021).

[48] R. –X. Zhang and S. D. Sarma, Anomalous Floquet Chiral Topological Superconductivity in a Topological Insulator Sandwich Structure, Phys. Rev. Lett. 127, 067001 (2021).

[49] H. Dehghani, M. Hafezi and P. Ghaemi, Light-induced topological superconductivity via Floquet interaction engineering, Phys. Rev. Research 3, 023039 (2021).

[50] S. Kitamura and H. Aoki, Floquet topological superconductivity induced by chiral many-body interaction, Commun Phys 5, 174 (2022).

[51] R. Ge, W. Broer and T. C. H. Liew, Floquet topological polaritons in semiconductor microcavities, Phys. Rev. B 97, 195305 (2018).

[52] W. Zheng and H. Zhai, Floquet topological states in shaking optical lattices, Phys. Rev. A 89, 061603(R) (2014).

[53] S. Choudhury and E. J. Mueller, Stability of a Floquet Bose-Einstein condensate in a one-dimensional optical lattice, Phys. Rev. A 90, 013621 – Published 21 July 2014.

[54] T. A. Sedrakyan, V. M. Galitski and A. Kamenev, Statistical Transmutation in Floquet Driven Optical Lattices, Phys. Rev. Lett. 115, 195301 (2015).

[55] C. Sträter, S. C. L. Srivastava and André Eckardt, Floquet Realization and Signatures of One-Dimensional Anyons in an Optical Lattice, Phys. Rev. Lett. 117, 205303 (2016).

[56] J. –Y. Shan, M. Ye, H. Chu, S. Lee, J. –G. Park, L. Balents and D. Hsieh, Giant modulation of optical nonlinearity by Floquet engineering, Nature 600, 235 (2021).

[57] M. –J. Yin, X. –T. Lu, T. Li, J.-J. Xia, T. Wang, X, -F, Zhang, and H. Chang, Floquet Engineering Hz-Level Rabi Spectra in Shallow Optical Lattice Clock, Phys. Rev. Lett. 128, 073603 (2022).

[58] Y. –G. Peng, C. –Z. Qin, D. –G. Zhao, Y. –X. Shen, X. –Y. Xu, M. Bao, H. Jia and X. –F. Zhu, Experimental demonstration of anomalous Floquet topological insulator for sound, Nat Commun 7, 13368 (2016).

[59] Z. Cheng, R. W. Bomantara, H. Xue, W. Zhu, J. Gong and B. Zhang, Observation of π/2 Modes in an Acoustic Floquet System, Phys. Rev. Lett. 129, 254301 (2022).

[60] S. Imhof, C. Berger, F. Bayer, J. Brehm, L. W. Molenkamp, T. Kiessling, F. Schindler, C. H. Lee, M. Greiter, T. Neupert and R. Thomale, Topolectrical-circuit realization of topological corner modes, Nat. Phys. 14, 925 (2018).

[61] R. Yu, Y. X. Zhao and A. P. A Schnyder, genuine realization of the spinless 4D topological insulator by electric circuits. Natl. Sci. Rev. 7, 1288 (2020).

[62] W. Zhang, D. Zou, W. He, J. Bao, Q. Pei, H. Sun and X. Zhang, Topolectrical-circuit realization of 4D hexadecapole insulator. Phys. Rev. B. 102, 100102(R) 2020.

[63] R. Li, B. Lv, H. Tao, J. Shi, Y. Chong, B. Zhang and H. Chen, Ideal type-II Weyl points in topological circuits. Natl. Sci. Rev. nwaa192 (2020).

[64] D. –A. Galeano, X. –X. Zhang, J. Mahecha, Topological circuit of a versatile non-Hermitian quantum system, arXiv: 2204. 01833 (2022).

[65] S. Weidemann, M. Kremer, S. Longhi, A. Szameit, Topological triple phase transition in non-Hermitian Floquet quasicrystals, Nature. 601, 354–359 (2022).

[66] A. Nagulu, X. Ni, A. Kord, M. Tymchenko, S. Garikapati, A. Alù, H. Krishnaswamy, Chip-scale Floquet topological insulators for 5G wireless systems, Nat Electron. 5, 300–309 (2022).

[67] S. S. Dabiri, H. Cheraghchi, Electric circuit simulation of Floquet topological insulators in Fourier space, J. Appl. Phys. 134, 084303 (2023).

[68] A. Eckardt, Colloquium: Atomic quantum gases in periodically driven optical lattices, Rev. Mod. Phys. 89, 011004 (2017).

[69] G. Salerno, T. Ozawa, H. M. Price, I. Carusotto, Floquet topological system based on frequency-modulated classical coupled harmonic oscillators. Phys. Rev. B. 93, 085105 (2016).
[70] L. Yuan, Q. Lin, M. Xiao, S. Fan, Synthetic dimension in photonics. Optica. 5, 001396 (2018).